\documentclass[letterpaper]{aa}

\usepackage{txfonts}
\usepackage{epsf}
\usepackage{amssymb}
\usepackage{bm}
\usepackage{natbib}

\newcommand{\req}[1]{Eq.~(\ref{#1})}

\newcommand{\etal}{et al.}
\newcommand{\Tb}{T_\mathrm{b}}
\newcommand{\Ts}{T_\mathrm{s}}
\newcommand{\barTs}{\overline{T}_\mathrm{s}}
\newcommand{\gcc}{\mbox{g cm$^{-3}$}}

\begin{document}
\headnote{Research Note}
\title{The magnetic structure of neutron stars and their
       surface-to-core temperature relation}
\titlerunning{Magnetic and thermal structure of neutron stars}
\author{
A. Y. Potekhin\inst{1,2,3}
\and
V. Urpin\inst{2,3}
\and
G. Chabrier\inst{1}
}
\authorrunning{A. Y. Potekhin et al.}

\institute{
    Ecole Normale Sup\'erieure de Lyon
    (C.R.A.L., UMR CNRS No.\ 5574),
    46 all\'ee d'Italie, 69364 Lyon Cedex 07, France
    \and
Ioffe Physico-Technical Institute,
    Politekhnicheskaya 26, 194021 St.~Petersburg, Russia
    \and
Isaac Newton Institute of Chile, St.~Petersburg Branch, Russia}

\offprints{A. Y. Potekhin,
    \email{palex@astro.ioffe.ru}}

\date{Received 14 June 2005 / Accepted 9 August 2005 }

\abstract{We study the relation between the mean effective surface 
temperature $\barTs$ and the internal temperature $\Tb$ for magnetic
neutron stars, assuming that the magnetic field near the surface has 
a presumably small-scale structure. The heavy-element (iron) and 
light-element (accreted) heat-blanketing envelopes are considered, 
and the results are compared with the case of a dipole magnetic field. 
We argue that the difference in the $\Tb(\barTs)$-relation for 
different magnetic configurations is always much smaller than a 
possible difference caused by variations of the chemical composition 
in the envelope.
}

\maketitle

\section{Introduction}

The most direct evidence for the nature of a poorly known 
superdense matter in neutron stars is likely to emerge from 
comparisons of surface thermal radiation measurements with 
predictions of neutron star cooling models (e.g., 
\citealt{pethick92,page97,page98,YakovlevPethick}). 
Modeling of neutron
star cooling is a complex problem that, generally, requires 
calculations of the temperature profile from the surface to the 
core at various stages of the evolution. The temperature profile 
for nonmagnetic neutron stars has been the subject of study for
many authors (e.g., \citealt{GPE,NomotoTsuruta,PCY},
and references therein). 
The key issue of these 
studies is the so-called $\Tb(\Ts)$-relation used in cooling 
models ($\Ts$ and $\Tb$ are the surface and internal 
temperature, respectively).

Most of the neutron stars, however, possess surface magnetic 
fields $B \sim 10^{12}- 10^{13}$ G, and some neutron stars are 
possibly magnetars with $B \gtrsim 10^{14}$ G. The internal 
magnetic field can be even higher. Such strong fields can 
affect the properties of plasma in neutron stars and
alter the $\Tb(\Ts)$-relation. 
In general, the magnetic field strength and direction varies over the
stellar surface, and hence $\Ts$ may be different
for different surface points. In many applications (e.g., in the
neutron-star cooling theory) it is sufficient to know 
the \emph{mean effective} surface temperature $\barTs$ instead of
the position-dependent $\Ts$.
The effective temperature is defined by the Stefan law,
$
   L = 4\pi R^2\sigma \barTs^4,
$
where $\sigma$ is the Stefan--Boltzmann constant, 
and $L$ is the thermal luminosity in a local
neutron-star reference frame, integrated over the surface. 
The apparent luminosity
measured by a distant observer is
$L^\infty=(1-r_g/R) \, L$,
and the apparent surface temperature inferred
by the observer from the radiation spectrum is
$\Ts^\infty=\Ts\,\sqrt{1-r_g/R}$
(e.g., \citealt{Thorne}).

The effects of a strong 
magnetic field on thermodynamic and kinetic properties of the 
outer neutron star layers have been reviewed, for instance, by 
\citet{YaK} and \citet{elounda}. 
The thermal structure of magnetized neutron stars has been
analyzed by a number of authors, often 
adopting a simplified magnetic 
geometry. For instance, much attention has been paid to the cases
of the radial magnetic field (e.g., 
\citealt{hern85,kvr88,schaaf90,hh-theory}) or 
the tangential field (with the field lines parallel to the 
surface; e.g.,
\citealt{hern85,schaaf90,hh-theory}).
The quantizing radial field field decreases 
the difference between $\Ts$ and $\Tb$, whereas
the tangential field increases this difference.

The case of an arbitrary inclination of the field lines has been 
considered by \citet{GH83} and \citet{page95} who 
argued that the $\Tb(\Ts)$-relation for such a field can be
constructed in a simple way from the relations for the radial 
and tangential magnetic fields. Numerical calculations
of the thermal structure \citep{PY01,PYCG} confirm an 
accuracy of this approximation.
These calculations show that the
$\Tb(\barTs)$-relation for the dipole magnetic 
configuration is almost independent of the field 
for $B \lesssim\mbox{a few}\times10^{13}$ G. 
The reason of such behaviour is the compensation of a decrease 
in the thermal flux near the magnetic equator  by an increase in the flux 
near the pole.
 
This result can essentially simplify the cooling 
calculations for magnetized neutron stars 
with dipole magnetic fields. It is likely, however, that the 
surface magnetic field of neutron stars departs from a  
dipole configuration. 
\citet{AronsSch} and 
\citet{Arons93,Arons00} noted that pulsars with long periods 
require a more complex field configuration than a dipole if pair 
creation is essential for the mechanism of radio emission. 
\citet{GilMitra} and \citet{GilMelikidze} also noted that radio 
emission of many pulsars
can be explained if one adopts 
the model with a strong ($\gtrsim 10^{13}$ G) and complex surface field 
with a small curvature of the field lines ($\sim 10^{5}$ cm).
\citet{GilSendyk} found that the behaviour 
of drifting subpulses observed in many pulsars is 
consistent with the vacuum gap maintained by a strong sunspot-like 
magnetic field. 
Recent observations of the X-ray spectra 
of some pulsars provide an opportunity to estimate the 
magnetic field near the neutron star surface. For example, a possible 
interpretation of the feature observed in the spectrum of PSR B1821--24
as a cyclotron emission (\citealt{psrB1821-24}; see, however,
\citealt{Mineo-ea04}) indicates
that the local magnetic fields on the neutron 
star surface can exceed the conventional dipole field
inferred from the spin-down data.

The growing evidence of the distinction between the
local field strength at the stellar surface and the global
dipole field suggests that this can be a general phenomenon in
neutron stars. Several theoretical explanations of this
phenomenon were suggested.
\citet{Ruderman91} considered a ``plate tectonics'' model,
in which a complex configuration of the surface magnetic field
of a neutron star is a result of neutron-star crust cracking
and platelet movements.
\citet{RuZhuChen} found that this model agrees with 
observations of pulsar glitches.
\citet{Geppert-ea03} suggested a Hall-driven mechanism of
the formation of the sunspot-like magnetic-field structure
at the neutron-star surface.
The presence of 
small scale field components can also be 
generated during the initial convective stage of evolution, e.g.,
by the turbulent mean-field dynamo
\citep{ThompsonDuncan93,Bonanno-ea03,Bonanno-ea05}. 
The unstable stage in proto-neutron stars lasts
$\sim 30$--40 s, which is sufficient for dynamo to reach a 
saturation level. 
The magnetic field generated 
in proto-neutron stars will be frozen into the 
crust that starts to form almost immediately after convection 
stops. Since the crustal conductivity is high, both the large and 
relatively small scale ($\sim 10^{5}$ 
cm) fields can survive during a long time comparable to 
the lifetime of pulsars, $\sim 10^{8}$ yr \citep{UrpinGil04}.

If the dipole field in neutron stars is accompanied by a stronger 
small-scale field, then the thermal structure of
the surface layers can be changed qualitatively. 
For instance, the difference between the equator 
and polar temperature should be much reduced. 
The small-scale field can also affect the
total photon flux at given $\Tb$ 
-- that is the $\Tb(\barTs)$ relation.
In this paper, we consider the thermal structure 
and calculate the $\Tb(\barTs)$ relation in the case of a neutron star
with a strong small-scale magnetic field in the envelope.
 
\section{Statement of the problem}

The thermal structure of the neutron star envelope is considered in
steady state plane-parallel approximation. 
We assume that $\Tb$ is equal to the temperature at the 
inner boundary of the envelope (taken at the neutron drip 
density $4 \times 10^{11}$ \gcc) and does not
vary over this boundary.
   
The statement of the problem for calculations of the temperature
profile in neutron star envelopes has been described in detail by
\citet{PYCG}. In the present study, the only difference
is the expression for the heat flux, that should
incorporate the effect of small-scale  magnetic fields. In the 
magnetic field $\vec{B}$, the heat flux is related to the 
temperature gradient by
\begin{equation}
\vec{F} = - \kappa_{\parallel}(B) \nabla_{\parallel} T - 
\kappa_{\perp}(B)
\nabla_{\perp} T - \kappa_{\wedge}(B) \vec{b} \times \nabla T,
\label{flux}
\end{equation}  
where the tensor and vector components along and across the field 
are marked by $\parallel$ and $\perp$, respectively, and 
$\kappa_{\wedge}$ is the so-called Hall component; $\vec{b} = 
\vec{B}/B$. Generally, all components of the thermal conductivity
tensor depend on the field strength, $B$. The vector components
in Eq.~(\ref{flux}) are
\begin{equation}
\nabla_{\parallel} T = \vec{b} \cdot(\vec{b}\cdot \nabla T) , \;\;\;
\nabla_{\perp} T = \nabla T - \vec{b}\cdot (\vec{b}\cdot \nabla T).
\end{equation}

We assume that small-scale fields are stronger 
than the large-scale field (as it follows from the turbulent 
dynamo models, e.g., \citealt{ThompsonDuncan93,Bonanno-ea05}),
and that the
length-scale of a temperature variation over the surface is much
larger than the vertical length-scale. 
This allows us to apply the local plane-parallel approximation for 
studying the mean thermal structure of the heat-blanketing envelope
(see \citealt{GPE} and \citealt{PYCG}, for discussion).
Then, averaging Eq.~(\ref{flux}) over
the directions of the magnetic field (that is approximately equivalent
to averaging over the stellar surface), we obtain for the mean heat
flux
\begin{equation}
\langle F \rangle = \frac{1}{3} [\kappa_{\parallel}(B) + 
2 \kappa_{\perp}(B)]
\frac{d T}{d z}, 
\label{dTdz}
\end{equation} 
where $z$ is the local proper depth 
in the envelope. This is equivalent to Eq.~(7) of \citet{PYCG},
but with the effective thermal conductivity $\kappa$
replaced by its average over all angles. 
Thus, unlike \citet{PYCG}, where the variations in strength and
direction of the magnetic field were assumed
smooth over the stellar surface and
small compared to the temperature gradient in the heat-blanketing 
envelope, which allowed us to treat the field as
locally constant, here we consider the opposite limit
of a rapidly varying field direction (i.e., ``entangled''
field lines).
For 
simplicity, we assume in what follows that the average 
strength of small scale magnetic fields does not depend on the 
depth.

\begin{figure*}
\centering
\epsfxsize=\textwidth
\epsffile{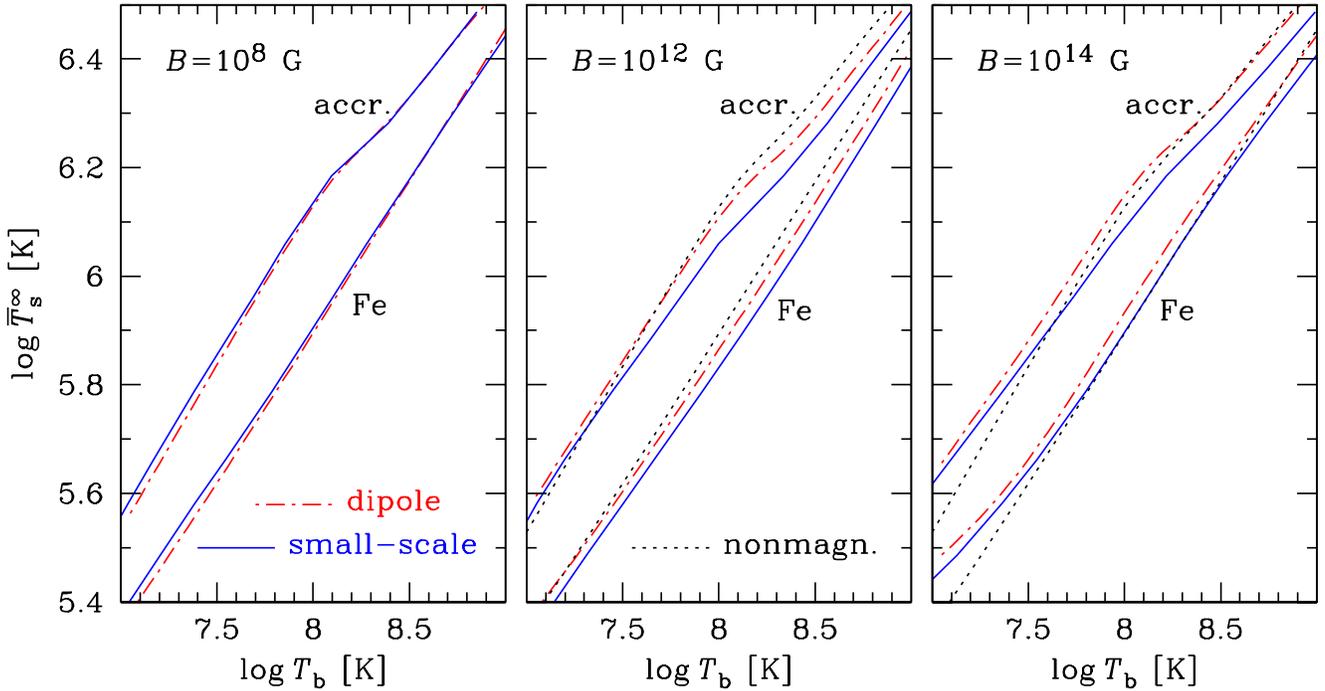}
\caption{The redshifted effective surface temperature
versus the internal temperature for a canonical neutron star 
($M=1.4\,M_\odot$, $R=10$ km) with the magnetic field $B=10^8$, 
$10^{12}$, and $10^{14}$~G (left, middle, and right panels)
and for the case of dipole (dot-dash lines) and small-scale (solid 
lines) fields.
The upper and lower curves represent the accreted and iron
heat-blanketing envelope, respectively.}
\label{fig:tstb}
\end{figure*}

We calculate the thermal structure of a neutron star envelope by 
integrating \req{dTdz} with the use of the numerical
scheme developed previously \citep{PCY,PYCG}.
We examine models of envelopes composed of iron
and accreted material. The accreted envelope is assumed to
be the same as described by \citet{PCY}:
the surface H layer is followed by He, C, O, and Fe layers.
The physics input
(equation of state, radiative opacities, and electron
thermal conductivities) is described in \citet{PYCG}. 
Note some uncertainty of this input for the outermost layer 
of the iron envelope  
where Fe plasma is only partly ionized. In this case,
the equation of state and opacities are based
on the Thomas-Fermi and mean-ion approximations.
For the accreted envelope, in contrast, we use
the accurate equation of state and opacities 
for the partially ionized hydrogen plasma in a strong
magnetic field (\citealt{PC04} and references therein).

\section{Results}

The calculated temperature profiles and their dependences on $B$
and $\Tb$ are quite similar to those presented in \citet{PYCG}.
Therefore we will not describe the results in detail but focus on the 
comparison of the $\Tb(\barTs)$ relation for the small-scale and
dipole fields.

\begin{figure}
\centering
\epsfxsize=86mm
\epsffile{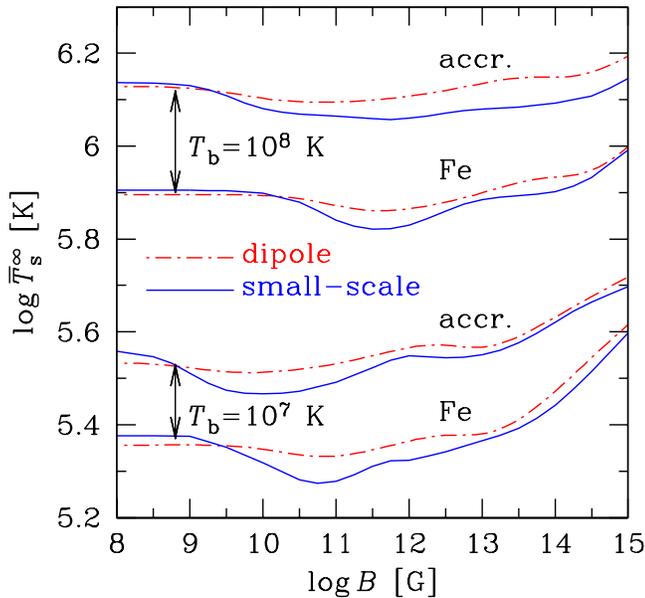}
\caption{The dependence of the mean effective temperature on $B$ 
for the dipole (dot-dash lines) and small-scale (solid lines) 
fields. The curves are shown for the accreted and iron envelopes 
and the internal temperature $\Tb=10^7$~K and $10^8$~K (marked 
near the curves).}
\label{fig:ts_b}
\end{figure}

The examples of such a comparison are shown 
in Figs.~\ref{fig:tstb} and~\ref{fig:ts_b}. 
The panels (from left to right)
in Fig.~\ref{fig:tstb} 
correspond to $B=10^8$~G, $10^{12}$~G, and $10^{14}$~G, 
typical for millisecond pulsars, ordinary radio pulsars, and
anomalous X-ray pulsars, respectively; $B$ is the average field 
strength in the case of a small-scale field and the field at the 
magnetic pole in the dipole field model. For convenience, the 
non-magnetic $\Tb(\barTs)$-relations are shown in the middle and right
 panels of Fig.~\ref{fig:tstb} by the dotted lines.

Figure~\ref{fig:ts_b} shows the dependence of $\barTs^\infty$ 
on $B$ for different models and for two $\Tb$ values. 
Independent of the field geometry, a weak field ($B\lesssim10^8$~G) 
does 
not affect the $\Tb(\barTs)$-relation for both the iron 
and accreted envelopes. 
With increasing $B$, the effective temperature
 first decreases and then increases
nonmonotonically. This behaviour is explained as follows.
The value of $\Ts$ at a given $\Tb$ is controlled mainly
by the opacity values in the ``sensitivity strip'' \citep{GPE}
located typically at $\rho\sim10^4$--$10^8$ \gcc,
where heat transport is provided by the electron conduction.
In a strong magnetic field, the electron thermal conductivity 
as a function of $\rho$ undergoes oscillations when 
the electron Fermi energy crosses the magnetic Landau levels
 (see, e.g., \citealt{YaK,elounda}).
The first oscillation at 
$\rho=\rho_B=7045\,B_{12}^{3/2}\,(A/Z)$ \gcc\ is
the strongest one (here $B_{12}\equiv B/10^{12}$~G and
$A$ and $Z$ are the atomic weight and charge numbers);
the classical (non-oscillating) magnetic electron conductivity
is recovered at $\rho\gg\rho_B$. In Fig.~\ref{fig:ts_b},
a decrease of $\barTs^\infty$ first occurs in the classical
regime, where the effect of the magnetic field amounts to
a suppression of $\kappa_\perp$ in \req{dTdz}. 
With further increase of $B$, the field becomes quantizing
in the sensitivity strip, leading to an increase of $\barTs^\infty$.
The increasing pieces of curves in Fig.~\ref{fig:ts_b}
are wavy because of quantum oscillations
of $\kappa_\parallel$, which are smoothed (but not entirely)
by the integration of \req{dTdz} over the sensitivity strip.

In the case of the dipole magnetic field,
the differences between the magnetic and 
nonmagnetic $\Tb(\barTs)$-relations become appreciable only in a 
very strong magnetic field, $B\gtrsim10^{14}$~G, and if the internal 
temperature is relatively low, $\Tb\lesssim3\times10^7$~K. In this case, 
$\barTs$ is noticeably increased by the magnetic field. 
The increase of a 
\emph{mean} effective temperature is however quite moderate in 
a comparison to the effect of such a strong field on the \emph{local} 
temperature value: a significant increase of $\Ts$ at the magnetic pole 
and a sharp decrease at the equator (see, e.g., \citealt{PY01,PYCG}). 
For lower magnetic field and higher temperature,
the difference in the magnetic geometry does not yield a considerable
departure in the $\Tb(\barTs)$-relation. 

The difference
in a magnetic configuration can manifest itself 
at 
$B\gtrsim10^{10}$~G. 
The small-scale field, as a rule, results in a lower $\barTs$
at a given $B$ compared to the dipole field. The difference in 
$\barTs$ for the small-scale and dipole fields depends generally 
on $\Tb$, $B$, and the mass of the accreted light-element material
(see Fig.~\ref{fig:ts_b}), but never exceeds 20\%. 
However, this difference in $\Ts$
can result in a factor of $\sim 2$ in the luminosity. In all
considered models, the uncertainty in a chemical composition of 
the envelope (light versus heavy elements) causes a much  
larger difference in $\barTs$ than the uncertainty in the magnetic 
field geometry (dipole versus small-scale).

To estimate the effect of a small-scale magnetic field on the 
$\Tb(\barTs)$-relation, one can use the fitting formulae derived for 
the dipole field (see the Appendix of \citealt{PYCG}) where 
$\cos^2\theta$ and $\sin^2\theta$ should be replaced 
by $1/3$ and $2/3$, respectively. Such fit reproduces the 
present numerical results with an accuracy $\leq 10$\%.

\section{Discussion}

We have considered the relation between the surface and internal
temperature in neutron stars in the case where small scale magnetic
fields near the surface are stronger than a large scale (e.g., 
dipole) field. Calculations show that the difference in the
$\Tb(\barTs)$-relation between the stars 
with small-scale and dipole fields is not very significant, although,
generally, it should be taken into account if highly accurate
thermal luminosity calculations are required.
This can be important for the interpretation of future
high-precision measurements of the neutron star thermal 
radiation.

For both the small-scale and dipole fields, the departure from the 
thermal structure of a nonmagnetic neutron star is relatively small,
if the typical field strength $B<10^{14}$~G. This is caused by the 
fact that the increase of thermal insulation near the region where 
the field lines are tangential is well compensated by the decrease
of the insulation in the region where the field is normal to the surface. 
As seen from Fig.~\ref{fig:ts_b}, for the dipole field this
compensation is generally more efficient than for the small-scale
field.

For a stronger field, an increase of the thermal conductivity
along the field lines due to the magnetic quantization effects
turns out to be so strong that it cannot be fully compensated
by a decrease of the conductivity in the perpendicular direction.

Our results indicate that the geometry and strength of 
the magnetic field are likely unimportant for the average thermal 
structure of neutron stars, if the field strength is moderate
($B \lesssim 10^{14})$. Therefore, the magnetic field of a moderate 
strength can affect the neutron star cooling via Joule heating 
rather than via the $\Tb(\barTs)$-relation. Joule heating is of 
particular importance at the late evolutionary stage and can 
maintain a relatively high surface temperature $\gtrsim 10^{5}$~K
during a very long time comparable to the decay time of the 
magnetic field \citep{Miralles-ea98}.

\begin{acknowledgements}
The work of A.Y.P.\ and G.C.\ was partially supported
by the CNRS French-Russian grant PICS 3202.
The work of A.Y.P.\ was also supported in part by
the RLSS grant 1115.2003.2 
and the RFBR grants 05-02-16245 and 03-07-90200. 
\end{acknowledgements}


\begin{thebibliography}{}


\bibitem[Arons(1993)]{Arons93}
Arons, J. 1993,
ApJ, 408, 160

\bibitem[Arons(2000)]{Arons00}
Arons, J. 2000,
in Pulsar Astronomy --- 2000 and beyond, Proc. IAU Coll.\ 177,
 ed.\ M.~Kramer, N.~Wex, \& R.~Wielebinski, ASP Conf.\ Ser., vol.\,202
(San Francisco: ASP), 449

\bibitem[Arons \& Scharlemann(1979)]{AronsSch}
Arons, J., \& Scharlemann, E. T. 1979,
ApJ, 231, 854

\bibitem[Becker et al.(2003)]{psrB1821-24}
Becker, W., Swartz, D. A., Pavlov, G.G.,  et al. 2003,
ApJ, 594, 798

\bibitem[Bonanno et al.(2003)Bonanno, Rezzolla, \& Urpin]{Bonanno-ea03}
Bonanno, A., Rezzolla, L., \& Urpin, V. 2003,
A\&A, 410, L33

\bibitem[Bonanno et al.(2005)Bonanno, Urpin, \& Belvedere]{Bonanno-ea05}
Bonanno, A., Urpin, V., \& Belvedere, G. 2005,
A\&A (in press)



\bibitem[Geppert et al.(2003)]{Geppert-ea03}
Geppert, U., Rheinhardt, M., \& Gil, J.	2003,
A\&A, 412, L33

\bibitem[Gil \& Melikidze(2002)]{GilMelikidze}
Gil, J., \& Melikidze, G. 2002,
ApJ, 577, 909

\bibitem[Gil \& Mitra(2001)]{GilMitra}
Gil, J., \& Mitra, D. 2001,
ApJ, 550, 383

\bibitem[Gil \& Sendyk(2000)]{GilSendyk}
Gil, J., \& Sendyk, M. 2000,
ApJ, 541, 351

\bibitem[Greenstein \& Hartke(1983)]{GH83}
Greenstein, G., \& Hartke, G. J. 1983,
ApJ, 271, 283

\bibitem[Gudmundsson et al.(1983)Gudmundsson, Pethick, \& Epstein]{GPE}
Gudmundsson, E. H., Pethick, C. J., \& Epstein, R. I. 1983, 
ApJ, 272, 286
 
\bibitem[Hernquist(1985)]{hern85}
Hernquist, L. 1985, MNRAS, 213, 313
 
\bibitem[Heyl \& Hernquist(1998)]{hh-theory}
Heyl, J. S., \& Hernquist, L. 1998, MNRAS, 300, 599

\bibitem[Mineo et al.(2004)]{Mineo-ea04}
Mineo, T., Cusumano, G., Massaro, E., Becker, W., \& Nicastro, L. 2004,
A\&A, 423, 1045
 
\bibitem[Miralles et al.(1998)Miralles, Urpin, \& Konenkov]{Miralles-ea98}
Miralles, J., Urpin, V., \& Konenkov, D. 1998,
ApJ, 503, 368

\bibitem[Nomoto \& Tsuruta(1987)]{NomotoTsuruta}
Nomoto, K., \& Tsuruta, S. 1987,
ApJ, 312, 711

\bibitem[Page(1995)]{page95}
Page, D. 1995, ApJ, 442, 273

\bibitem[Page(1997)]{page97} 
Page, D. 1997, ApJ, 479, L43

\bibitem[Page(1998)]{page98}
Page, D. 1998, 
in 
The Many Faces of Neutron Stars, 
ed.\ R. Buccheri, J. van Paradijs, \& M. A. Alpar
(Dordrecht: Kluwer), 539


\bibitem[Pethick(1992)]{pethick92}
Pethick, C. J. 1992,
Rev.\ Mod.\ Phys., 64, 1133

\bibitem[Potekhin \& Chabrier(2004)]{PC04}
Potekhin, A. Y., \& Chabrier, G. 2004,
ApJ, 600, 317

\bibitem[Potekhin \& Yakovlev(2001)]{PY01}
Potekhin, A. Y., \& Yakovlev, D. G. 2001,
A\&A, 374, 213

\bibitem[Potekhin \etal(1997)Potekhin, Chabrier, \& Yakovlev]{PCY}
Potekhin, A. Y., Chabrier, G., \& Yakovlev, D. G. 1997,
A\&A, 323, 415

\bibitem[Potekhin et al.(2003)]{PYCG}
Potekhin, A. Y., Yakovlev, D. G., Chabrier, G., \& Gnedin, O.Y. 2003,
ApJ, 594, 404

\bibitem[Ruderman(1991)]{Ruderman91}
Ruderman, M., 1991,
ApJ, 382, 587

\bibitem[Ruderman et al.(1998)Ruderman, Zhu, \& Chen]{RuZhuChen}
Ruderman, M., Zhu, T., \& Chen, K. 1998,
ApJ, 492, 267


\bibitem[Schaaf(1990)]{schaaf90}
Schaaf, M. E. 1990, A\&A, 227, 61


\bibitem[Thompson \& Duncan(1993)]{ThompsonDuncan93}
Thompson, C., \& Duncan, R. C. 1993,
ApJ, 408, 194

\bibitem[Thorne(1977)]{Thorne}
Thorne K. S., 1977, ApJ 212, 825

\bibitem[Urpin \& Gil(2004)]{UrpinGil04}
Urpin, V., \& Gil, J. 2004, A\&A, 415, 305

\bibitem[Van Riper(1988)]{kvr88}
Van Riper, K. A. 1988,
ApJ, 329, 339

\bibitem[Ventura \& Potekhin(2001)]{elounda}
Ventura, J., \& Potekhin, A. Y. 2001,
in NATO Science Ser. C, 567,
The Neutron Star -- Black Hole Connection,
ed.\ C. Kouveliotou, J. Ventura, \& E. P. J. van den Heuvel
(Dordrecht: Kluwer), 393 

\bibitem[Yakovlev \& Kaminker(1994)]{YaK}
Yakovlev, D. G., \& Kaminker, A. D. 1994,
in The Equation of State in Astrophysics,
ed.\ G. Chabrier \& E. Schatzman
(Cambridge: Cambridge University Press), 214

\bibitem[Yakovlev \& Pethick(2004)]{YakovlevPethick}
Yakovlev, D.G., \& Pethick, C.J. 2004,
ARA\&A, 42, 169--210.

\end{thebibliography}
\end{document}